\documentclass[9pt,twocolumn,twoside]{osajnl}

\usepackage{amsmath}

\usepackage{epsfig}
\usepackage{graphicx}
\usepackage{epstopdf}
\usepackage{amssymb}
\usepackage{color,soul}

\newcommand{\bs}{\boldsymbol}

\journal{ol} 

\setboolean{shortarticle}{true}

\title{
On z-coherence in self-focusing
}

\author[1]{F. Gori}
\author[2]{R. Mart\'inez-Herrero}
\author[2]{G. Piquero}
\author[3]{J. C. G. de Sande}
\author[4]{O. Korotkova}
\author[1,*]{M. Santarsiero}

\affil[1]{Dipartimento di Ingegneria Industriale, Elettronica e Meccanica, Universit\`a Roma Tre, Via V. Volterra 62, 00146 Rome, Italy}
\affil[2]{Departamento de \'{O}ptica, Universidad Complutense de Madrid, Ciudad Universitaria,  28040 Madrid, Spain}
\affil[3]{ETSIS de Telecomunicaci\'{o}n, Universidad Polit\'{e}cnica de Madrid, Campus Sur 28031 Madrid, Spain}
\affil[4]{Department of Physics, University of Miami, Coral Gables, FL, USA}
\affil[*]{Corresponding author: msantarsiero@uniroma3.it
\newline
\newline
\bf © 2022 Optica Publishing Group. One print or electronic copy may be made for personal use only. Systematic reproduction and distribution, duplication of any material in this paper for a fee or for commercial purposes, or modifications of the content of this paper are prohibited.
\newline
\url{https://doi.org/10.1364/OL.455449}
}

\begin{abstract}
Both the intensity distribution and the degree of coherence between pairs of points along the propagation axis (z-coherence) are derived in closed form for a phenomenon of self-focusing produced by circularly coherent light. 
The first confirms results previously obtained numerically, while the second exhibits new complex features. The physical interpretation is obtained by a suitable pseudo-modal expansion that suggests an analogy with a simple two-mode structure. 
\end{abstract}

\setboolean{displaycopyright}{true}

\begin{document}             

\maketitle     

The self-focusing effect of partially coherent light upon free propagation has received much attention in recent years~\cite{Lajunen:OL11, Tong:JOSAA12, Mei:Ol14, Ding:JOSAA17, WuCai:OLT18, Sande:OE19, RMH:AS19,YuJ:OL20, Mei:OL21}, also for their potential applications in several fields, such as free space optical communications, particle trapping and manipulation. In particular  the intensity along the mean propagation axis ($z$-axis) was considered.  Other significant elements of the phenomenon are afforded, as we shall see, by the study of the pertinent coherence properties along the $z$-axis. They are generally meant to be a manifestation of the so-called {\em longitudinal} or {\em temporal} coherence, which depends on the light power spectrum through the Wiener-Kintchine theorem~\cite{ManWolf95}. The term longitudinal, however, is also used to denote correlation along the $z$-axis in the space-frequency approach, in which light correlations are expressed through the cross-spectral density (CSD) \cite{ManWolf95}. The latter refers to a specific temporal frequency and then to the monochromatic regime. To avoid confusion between denominations we shall use the name $z$-{\em coherence}  for correlations described by the CSD along the $z$-axis. 

We begin our analysis by recalling the propagation law of the CSD, $W(\bs r_1, z_1; \bs r_2, z_2)$, where $\bs r_1, \bs r_2$ are transverse position vectors orthogonal to the $z$-axis
\begin{equation}
\begin{array}{c}
\displaystyle W(\bs r_1, z_1; \bs r_2, z_2) =
\\
\displaystyle \int \int W_0(\bs \rho_1,\bs \rho_2) K_{z_1}^*(\bs r_1- \bs\rho_1) K_{z_2}(\bs r_2-\bs\rho_2) {\rm d}^2\rho_1 {\rm d}^2\rho_2 
\; ,
\end{array}
\label{gen01}
\end{equation}
$K_{z_j}$  denoting the free propagator from $z=0$ to $z=z_j$  $(j=1,2)$ \cite{ManWolf95} and $W_0(\bs \rho_1,\bs \rho_2)$ the CSD across the source. The CSD along the $z$-axis ($ r_1 =  r_2 = 0$) will be denoted by $W_z(z_1,z_2)$, which, in the paraxial approximation, is given by
\begin{equation}
\begin{array}{c}
\displaystyle W_z(z_1, z_2) = \frac{e^{{\rm i} k(z_2 - z_1)}}{\lambda^2 z_1 z_2} \times  
\\
\displaystyle \int_0^{\infty}{\rm d}\rho_1 \int_0^{\infty} {\rm d}\rho_2 \int_0^{2\pi}{\rm d}\phi_1 \int_0^{2\pi}{\rm d}\phi_2\, W_0(\bs \rho_1,\bs \rho_2) \times   
\\
\displaystyle   \rho_1 \rho_2 \exp\left[\frac{{\rm i}k}{2}\left(\frac{\rho_2^2}{z_2}-\frac{\rho_1^2}{z_1}  \right)  \right] 
\; ,
\end{array}
\label{gen02}
\end{equation}
where $\lambda$ denotes the wavelength of the radiation and $k=2 \pi/\lambda$. In the limiting  case of spatially incoherent radiation Eq. (\ref{gen02}) can be read  \cite{Rosen:OC95} as an analogous of the van Cittert-Zernike theorem \cite{ManWolf95}. 
Correlation behavior along the $z$-axis can be thought of as representative of what takes place along off-axis lines crossing the transverse correlation area.

We now consider the case of circular coherence \cite{Santarsiero:OL17} in which $W_0$ does not depend on the angular coordinates.
We assume the source to be of Gaussian profile with the following CSD
\begin{equation}
\begin{array}{c}
\displaystyle W_0(\rho_1,\rho_2)\propto \exp[-(\rho_1^2+\rho_2^2)/w_0^2] \,\mu(\rho_1, \rho_2)
\; ,
\end{array}
\label{sou01}
\end{equation}
where $w_0$ is a real constant and $\mu$ is the degree of coherence (DOC) across the source \cite{ManWolf95}. Exploiting the circular symmetry the propagated CSD takes on the form
\begin{equation}
\begin{array}{c}
\displaystyle W_z(z_1,z_2)=\frac{k^2 e^{{\rm i}k(z_2 - z_1)}}{z_1z_2}  \times
\\
\displaystyle  \int_0^{\infty}\int_0^{\infty} \exp\left[-\left(\frac{\rho_1^2+\rho_2^2}{w_0^2} \right)\right] \mu(\rho_1, \rho_2)    \times
\\
\displaystyle \exp\left[\frac{{\rm i}k}{2}\left(\frac{\rho_2^2}{z_2} - \frac{\rho_1^2}{z_1}\right)\right]  \rho_1 \rho_2 \; {\rm d}\rho_1 {\rm d}\rho_2
\; .
\label{sou02}
\end{array}
\end{equation} 

 As for the degree of coherence we adopt the quartic exponential structure 
\begin{equation}
\begin{array}{c}
\displaystyle \mu(\rho_1, \rho_2) =\exp\left[-\frac{(\rho_1^2-\rho_2^2)^2}{\delta^4}\right]
\; ,
\label{sou03}
\end{array}
\end{equation} 
with real constant $\delta$, previously used in some papers \cite{Lajunen:OL11, Tong:JOSAA12, Mei:Ol14}. 
Clearly enough, Eq.~(\ref{sou03}) specifies a structured DOC~\cite{Korotkova:P21,Ponomarenko:PX21}.
In Refs.~\cite{Ding:JOSAA17,Santarsiero:OL17b} the exponential function in $\mu$ was replaced by a sinc, but  the same dependence from $\rho_2^2 - \rho_1^2$ was adopted. 
Different or more general dependences from $\rho_2^2 - \rho_1^2$ was also used~\cite{Santarsiero:OL17,WuCai:OLT18, Sande:ST20, YuJ:OL20, Mei:OL21}. 
Some numerical tests suggest that other sensible forms of the DOC would lead to results similar to those we are going to illustrate.
%

To solve the propagation integral (\ref{sou02}) we let
\begin{equation}
q = \left( \frac{w_0}{\delta}\right)^2 ;
\;\;\;
\displaystyle  \tau_j = \frac{\rho_j^2}{w_0^2} \; ; 
\;\;\;
\displaystyle \zeta_j=\frac{z_j}{z_R}=\frac{z_j \lambda}{\pi w_0^2} 
\; ;
\;\; (j=1,2)
\label{sou06b}
\end{equation} 
with $z_R=\pi w_0^2/\lambda$. 
Note, in particular, that the parameter $q$ accounts for the global coherence of the source, small values of $q$ denoting high coherence.
In such a way Eq.~(\ref{sou02}) gives a CSD of the form
\begin{equation}
\begin{array}{c}
\displaystyle W_{\zeta}(\zeta_1,\zeta_2) =
\frac{e^{2 {\rm i} (\zeta_2 - \zeta_1) z_R^2 /w_0^2}}{\zeta_1\zeta_2}  \;
\times
\\
\displaystyle  \int_0^{\infty}\int_0^{\infty} \exp\left[-\tau_1\sigma_1 - \tau_2\sigma_2^*\right]    
\; \displaystyle \exp\left[-q^2(\tau_2 - \tau_1)^2\right]    {\rm d}\tau_1 {\rm d}\tau_2
\; ,
\label{sou07c}
\end{array}
\end{equation} 
where the complex function $\sigma(\zeta)=1-{\rm i}/\zeta$ has been introduced and we put, for brevity, $\sigma_j=\sigma(\zeta_j)$ $(j=1,2)$. The CSD turns out to be
\begin{equation}
\begin{array}{c}
W_\zeta(\zeta_1,\zeta_2)= 
\displaystyle \frac{\sqrt{\pi}}{2q} 
\frac{e^{2 {\rm i} (\zeta_2 - \zeta_1) z_R^2 /w_0^2}}
{2 \zeta_1 \zeta_2 - {\rm i} (\zeta_2 - \zeta_1)}
\; \times 
\\
\\
\left\{
e^{\frac{\sigma^2(\zeta_1)}{4q^2}}
{\rm erfc}
\left[
\displaystyle\frac{\sigma(\zeta_1)}{2q} 
\right]
+ 
e^{\frac{{\sigma^{*}}^2(\zeta_2)}{4q^2}}
{\rm erfc}
\left[
\displaystyle\frac{\sigma^*(\zeta_2)}{2q}  
\right]
\right\}
\, ,
\label{sou08}
\end{array}
\end{equation} 
where erfc stands for the complementary error function~\cite{Gradshteyn65}.
It is seen that the solution of the propagation integral in Eq.~(\ref{sou08}) involves the twodimensional Laplace transform~\cite{Gradshteyn65} of the degree of coherence. It should be noticed that the occurrence of the Laplace transform as the proper mathematical tool to study free propagation of light is rather uncommon, but its use has already been at the basis of some recent results~\cite{Santarsiero:P21}.

From Eq.~(\ref{sou08}), in particular, the intensity is evaluated as
\begin{equation}
\begin{array}{c}
I(\zeta)=  
\displaystyle \frac{\sqrt{\pi}}{2 \, q \, \zeta^2} 
\; \Re\left[e^{\frac{\sigma^2(\zeta)}{4q^2}} \, {\rm erfc}\left[\displaystyle\frac{\sigma(\zeta)}{2q}  \right] \right]
\; ,
\label{sou08b}
\end{array}
\end{equation} 
with $ \Re$ denoting the real part.
To the best of our knowledge, this is the first closed-form expression of the intensity along the $z$-axis in a self-focusing phenomenon. As for the CSD in Eq.~(\ref{sou08}) we are not aware of any (analytical or numerical) previous derivation.
 
\begin{figure}[!ht]
	\centering
	\includegraphics[width=7 cm] {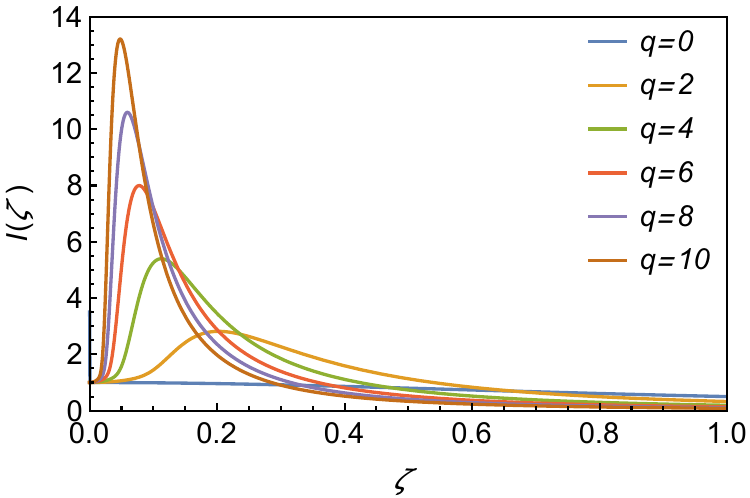}
	\caption{Axial intensity vs $\zeta$, according to Eq. (\ref{sou08b}), for values of $q$ increasing from 0 (coherent case) to 10 with step 2 (from lowest to highest peak).}
	\label{fig001}
\end{figure}

Plots of $I$ vs $\zeta$ for several values of $q$ are drawn in Fig.~\ref{fig001}. They agree with the general shapes previously seen through numerical methods \cite{Lajunen:OL11, Tong:JOSAA12, Mei:Ol14, Ding:JOSAA17, Sande:OE19, Sande:ST20}.

\begin{figure}[!ht]
\centering
\includegraphics[width=8 cm] {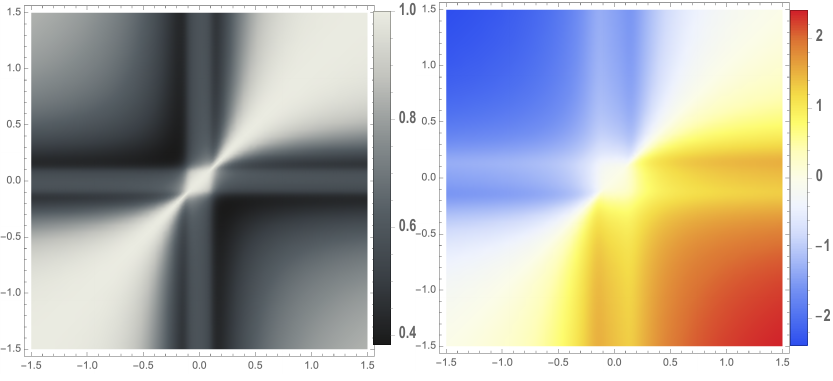}
\caption{Modulus  and argument  of $\mu$ across the plane ($\zeta_1$, $\zeta_2$)  for  $q=2$.}
\label{fig002}
\end{figure}

Let us now pass to the DOC. 
In dealing with z-coherence we shall consider also cases in which one of the two points has a negative $\zeta$. Indeed, the Fresnel diffraction integral can be also used in backpropagation.
2D plots of the modulus of $\mu(\zeta_1, \zeta_2)$ and its argument are shown in Fig.~\ref{fig002}  for $q=2$.
For a clearer visualization of the coherence between two axial points and the effect of the parameter $q$, Fig.~(\ref{fig003}) shows the modulus of $\mu$ for two points that are symmetrical with respect to $\zeta=0$ (i.e. $\zeta_1=-\zeta_2$), while Fig.~(\ref{fig004}) shows the same quantity when the distance between the two points is kept fixed (i.e. $\zeta_2=\zeta_1-\Delta \zeta$). It could be seen that, on increasing $q$, pairs of points for which the DOC becomes very low exist. 

{Another interesting aspect is the following. A quantitative analysis of Fig.~\ref{fig002} shows that, as the separation between $\zeta_1$ and $\zeta_2$ grows (for instance, keeping $\zeta_1$ fixed), the modulus of DOC does not generally tend to zero but to some positive value. This is reminiscent of what happens for two points in the far field (see \cite{ManWolf95}, Sec 5.2.1), in which case the modulus of DOC tends to one.  Here we show that something analogous holds when one of the two points is in the near field.
\begin{figure}[!ht]
\centering
\includegraphics[width=7 cm]{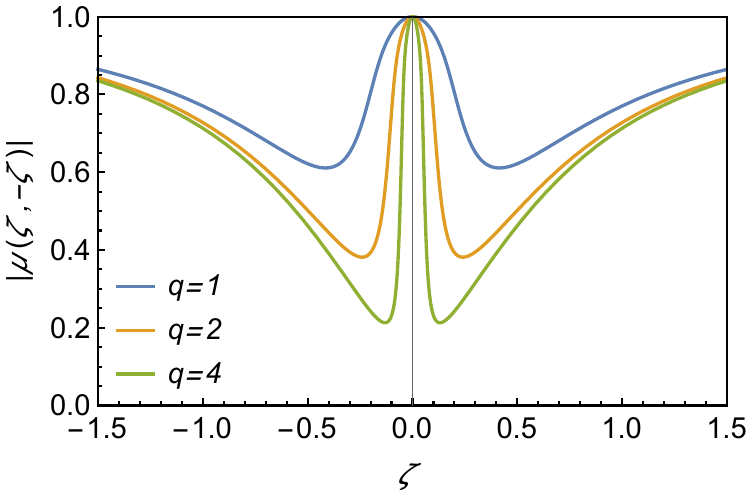}
\caption{Plot  of $|\mu(\zeta,-\zeta)|$ as a function of $\zeta$ for  $q=1,2,4$. }
\label{fig003}
\end{figure}
\begin{figure}[!ht]
\centering
\includegraphics[width=7 cm]{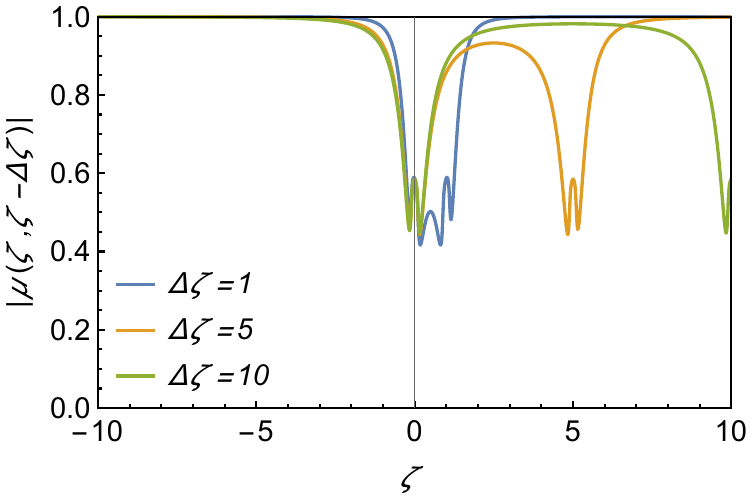}
\caption{Plot  of $|\mu(\zeta,\zeta-\Delta \zeta)|$ as a function of $\zeta$ for  $\Delta\zeta=1,5,10$ and $q=2$. }
\label{fig004}
\end{figure}

The behavior of the DOC as shown by the previous figures is rather complex. 
We now will show that a simplified model can be built with the help of a suitable set of pseudo-modes \cite{RMH:OL09}. 

A set of pseudo-modes generating a quartic DOC was constructed in the one-dimensional Cartesian case by Lajunen and Saastamoinen \cite{Lajunen:OL11} using the rules of Refs.~\cite{Gori:OL07,RMH:OL09,Gori:JOSAA21}. Such a structure, with its generalizations, has been used in various papers \cite{Tong:JOSAA12,Mei:Ol14,Ding:JOSAA17,Mei:OL21}. We remind it briefly. Starting from Eq. (11) of \cite{Gori:OL07}, namely 
\begin{equation}
\begin{array}{c}
\displaystyle W_0(\bs \rho_1,\bs \rho_2) = \int  p(\bs v) H^*(\bs \rho_1,\bs v) H(\bs \rho_2,\bs v) {\rm d}^2v
\; ,
\label{mod01}
\end{array}
\end{equation} 
where $H$ is an arbitrary kernel and $p$ is a non-negative function, which guarantees $W$ to be a non-negative definite kernel, we build a structure that leads to the CSD of Eqs.~(\ref{sou01}) and (\ref{sou03}). Explicitly, this is obtained with
\begin{equation}
\begin{array}{c}
\displaystyle H(\bs \rho,v) = \exp[-(b + {\rm i}v)\rho^2]
\; ,
\label{mod01b}
\end{array}
\end{equation} 
with constant $b>0$. Then Eq. (\ref{mod01}) becomes
\begin{equation}
\begin{array}{c}
\displaystyle W_0( \rho_1, \rho_2) = \int_{-\infty}^{\infty}  p(v) e^{-(b  - {\rm i}  v) \rho_1^2} e^{-(b + {\rm i} v)\rho_2^2}\, {\rm d}v =
\\
\displaystyle  e^{-b(\rho_1^2 + \rho_2^2)} \int_{-\infty}^{\infty}  p(v) e^{-   {\rm i}  v( \rho_2^2 - \rho_1^2)} \, {\rm d}v 
\; .
\label{mod02}
\end{array}
\end{equation} 

It is seen that $W_0$ is (up to proportionality factors) the Fourier transform of $p(v)$ in suitable units \cite{Gori:OL07}. This can specify a great number of significant DOCs depending on the non-negative function $p(v)$. In particular, if
\begin{equation}
\begin{array}{c}
\displaystyle p(v) = \exp(-a v^2)
\; ,
\label{mod02b}
\end{array}
\end{equation} 
with positive $a$, Eq. (\ref{mod02}) becomes
\begin{equation}
\begin{array}{c}
\displaystyle W_0( \rho_1, \rho_2) =\sqrt{\frac{\pi}{a}} \; e^{-b(\rho_1^2+\rho_2^2)} \; e^{- \frac{( \rho_1^2 - \rho_2^2)^2}{4a}   }
\; .
\label{mod02c}
\end{array}
\end{equation} 
which is functionally equivalent to Eqs. (\ref{sou01}) and (\ref{sou03}) with
$w_0^2=1/b$ and $\delta^4 = 4 a$.

The pseudo-mode structure $\exp[-(b + {\rm i} v)\rho^2]$ in its dependence from $b$ and $v$  is very simple. It  can be identified, up to a proportionality factor with that of a TEM$_{00}$ mode \cite{Siegman:Lasers86}. Such a remark shows that the CSD can be thought of as produced by a suitable superposition of uncorrelated Gaussian beams of the lowest order. All the beams have the same spot-size $w_0$ across the plane $z$= 0, which is independent of $v$. Each beam though has its own radius of curvature $R(v)$. 
More precisely, Eq. (\ref{mod01b}) implies
$v=-k/[2R(v)]$, so that $R(v)=-\pi/(\lambda v)$.

Except for the beam associated to $v=0$, all the Gaussian beams have, at $z=0$ a finite (positive or negative) radius of curvature. 
Accordingly, we have to determine the position, say $z_0(v)$ of the waist plane pertaining to the beam associated to a specific value of $v$, as well as the corresponding waist spot-size, say $u_0(v)$. This is done with the help of the  formulas for Gaussian beams \cite{Siegman:Lasers86}, which give
\begin{equation}
\begin{array}{c}
z_0(v) = \displaystyle\frac{\pi w_0^2}{\lambda}  \; \frac{ v w_0^2}{1+\left(v w_0^2  \right)^2}; 
\;\;\; \displaystyle u_0(v) = \frac{w_0}{\sqrt{1+\left(v w_0^2  \right)^2}}
\; .
\label{las05}
\end{array}
\end{equation} 
It can be seen that $|z_0(v)|$, starting from 0 at $v=0$, grows up to a maximum and then decreases toward 0, while the waist spot-size $u_0(v)$ has its maximum at $v=0$ and decreases on  both sides.

The complete expression of the Gaussian beam associated to a given value of $v$ is
\begin{equation}
\begin{array}{c}
\displaystyle f(r,z,v) =  \frac{Z(0,v)}{Z(z,v)}\exp{\left[\frac{{\rm i} k}{2Z(z,v)}r^2 \right]} 
\; ,
\label{las05b}
\end{array}
\end{equation} 
which is a solution of the paraxial wave equation where
\begin{equation}
\begin{array}{c}
\displaystyle  Z(z,v)=z - z_0(v) - {\rm i}  \frac{\pi u_0^2(v)}{\lambda }
\; .
\label{las06}
\end{array}
\end{equation} 
Using Eq. (\ref{las05}) $Z(z,v)$ can be written as
\begin{equation}
\begin{array}{c}
\displaystyle  Z(z,v) = z -  \frac{\pi w_0^2}{\lambda}\frac{1}{ v w_0^2 - {\rm i}}
= z - \displaystyle\frac{z_R}{ v w_0^2 - {\rm i}} 
\; ,
\label{las07}
\end{array}
\end{equation} 
and Eq.~(\ref{las05b}) gives, for the field along the $z$-axis,
\begin{equation}
\begin{array}{c}
\displaystyle f(0,z,v) = \displaystyle \frac{1}{1-(v w_0^2 - {\rm i})(z/z_R)} 
\, .
\label{las05c}
\end{array}
\end{equation} 

The effect of combining the infinite number of the just specified Gaussian beams is not intuitively evident but we can account for it by considering a simple model with only two uncorrelated modes~\cite{Sande:ST20,Visser:OC22}. 
Let us take two Gaussian modes, corresponding to two opposite values of $v$, say, $v_0$ and $-v_0$, with  $v_0>0$. From Eq.~(\ref{las05}) it follows that they have the same waist size, located at symmetrical positions with respect to the plane $z=0$. The corresponding $z$-dependent fields along the $z$-axis are
\begin{equation}
\begin{array}{c}
\displaystyle  V_\pm(\zeta) = \displaystyle \frac{1}{1-(\pm v_0 w_0^2 - {\rm i})  \zeta} 
\, .
\label{las09}
\end{array}
\end{equation} 
with, again, $\zeta=z/z_R$. Assuming that the two beams carry equal powers the CSD they produce is
\begin{equation}
\displaystyle  W_\zeta(\zeta_1,\zeta_2)=V_+^*(\zeta_1)V_+(\zeta_2) +V_-^*(\zeta_1)V_-(\zeta_2) 
\, .
\label{las10}
\end{equation} 

The on-axis intensity and DOC are easily evaluated. The former, normalized with respect to the intensity at $\zeta=0$, is shown in Fig.~\ref{fig005} for different values of $v_0 w_0^2$. Although not identical, as it is reasonable, the reported plots gives qualitatively account of the behavior of the analogous quantity shown in Fig.~\ref{fig001}.
\begin{figure}[!ht]
\centering
\includegraphics[width=7 cm]{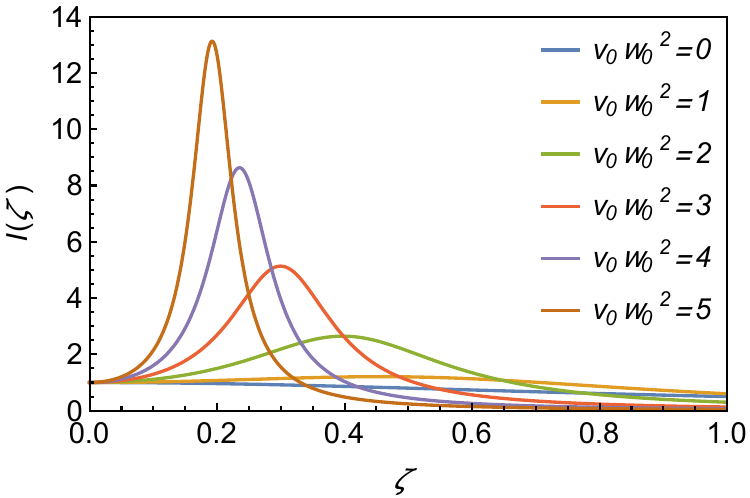}
\caption{Plot  of the on-axis intensity for the CSD in Eq.~(\ref{las10}), for values of $v_0 w_0^2$ from 0 (coherent case) to 5 with step 1 (from lowest to highest peak)}
\; .
\label{fig005}
\end{figure}

Analogous considerations can be done for the DOC. Figure~\ref{fig006} shows modulus and phase of the DOC across the plane ($\zeta_1,\zeta_2$) for $v_0 w_0^2=2$. On comparing it to Fig.~(\ref{fig002}), we note that the finest details have been lost in the model, but the main structure is still present. 
\begin{figure}[!ht]
\centering
\includegraphics[width=8 cm]{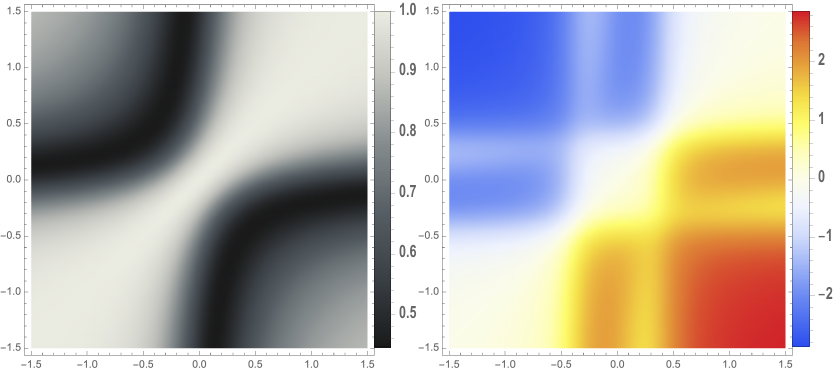}
\caption{Modulus  and argument of $\mu$ across the plane ($\zeta_1$, $\zeta_2$) for the CSD in Eq.~(\ref{las10}), with  $v_0 w_0^2=2$.}
\label{fig006}
\end{figure}
This is even more evident from Figs.~(\ref{fig007}) and (\ref{fig008}). In the first one the plot of  $|\mu|$ as a function of $\zeta_1$ is shown with $\zeta_2=-\zeta_1$, for $v_0 w_0^2=1,2,4$. In the second one the distance between the two points ($\Delta \zeta$) is kept fixed and $v_0 w_0^2=2$. The reported behaviors fairly reproduce the ones in Figs.~(\ref{fig003}) and (\ref{fig004}). 
\begin{figure}[!ht]
\centering
\includegraphics[width=7 cm]{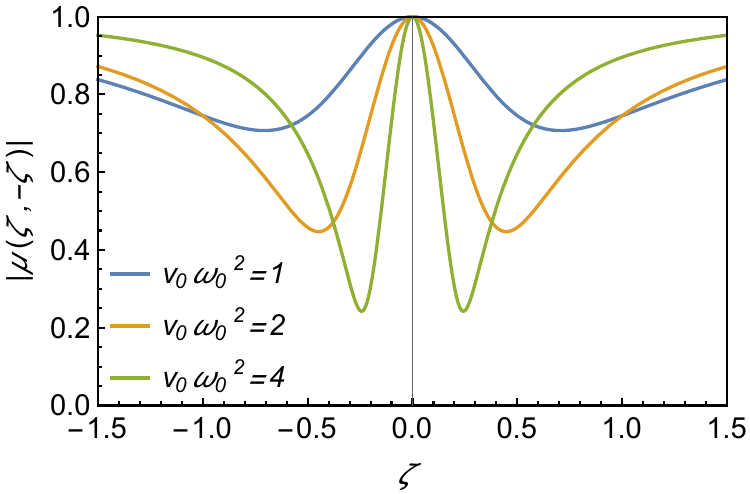}
\caption{Plot  of  $|\mu|$ for the CSD in Eq.~(\ref{las10}), as a function of $\zeta$ with $\zeta_2=-\zeta$, for $v_0 w_0^2=1,2,4$ }
\label{fig007}
\end{figure}
\begin{figure}[!ht]
\centering
\includegraphics[width=7 cm]{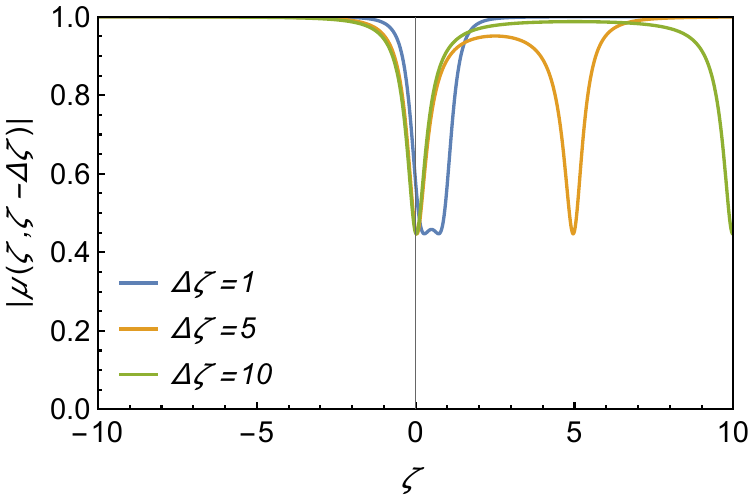}
\caption{Plot  of $|\mu(\zeta,\zeta-\Delta\zeta)|$ for the CSD in Eq.~(\ref{las10}), as a function of $\zeta$ for $v_0 w_0^2=2$, with $\Delta\zeta=1,5,10$}
\label{fig008}
\end{figure}
This simple model also provides a key to understanding the effect mentioned above about the limiting value of $|\mu|$ for increasing distance between the points. Let us consider, for example, $z_1=0$ and $z_2 \to \infty$. In this limit, as follows from Eq.~(\ref{las09}), the two beams have the same amplitude both at $z_1$ and at $z_2$, but they acquire different phases during propagation. In fact, although they travel the same distance, only one of them passes through its waist plane, so that its phase is reduced by the phase anomaly~\cite{Siegman:Lasers86}. The phase difference between the two modes causes the reduction of the DOC. On the other hand, when $z_2 \to \infty$, the phase difference stabilizes to a limiting value, and so does the DOC. Analytically, from Eq.~(\ref{las09}) it turns out that $|\mu| \to [(v_0 w_0^2)^2+1]^{-1/2}$.

In conclusion, the z-coherence properties of the field radiated by a partially coherent source giving rise to the self-focusing phenomenon have been studied, and the closed-form expression of the propagated CSD has been obtained for a commonly used source of this class. Significant features of the propagated intensity and of the degree of coherence can be reproduced using a simple source model, based on the superposition of only two mutually uncorrelated gaussian beams. 
It could be seen that our investigation method can be extended to hollow beams without vortexes.

\vspace{.3cm}
{\bf Funding.}  	 

\vspace{.3cm}
{\bf Disclosures.}  The authors declare no conflicts of interest.

\vspace{.3cm}
{\bf Data Availability.}  Data underlying the results presented in this paper are not publicly available at this time but may be obtained from the authors upon reasonable request.

\end{document}